\documentclass[titlepage]{amsart}

\usepackage{cite}

\usepackage[pdftex]{graphicx}
\usepackage[tight,footnotesize]{subfigure}
\usepackage{hyperref}
\usepackage{url}

\usepackage{amssymb}

\usepackage[usenames]{color}

\usepackage{algorithm}
\usepackage{algpseudocode}
\usepackage{paralist}
\usepackage{listings}
\usepackage{multirow}

 \usepackage[foot]{amsaddr}

\newcommand{\trih}{\mathcal{T}_{H}}
\newcommand{\trihr}{\mathcal{T}_{H}^{r}}
\newcommand{\fach}{\mathcal{E}_{H}}
\newcommand{\matr}[1]{\mathbf{\underline{#1}}}

\title[On the Implementation of a Scalable Simulator for MHM]{On the Implementation of a Scalable Simulator for Multiscale Hybrid-Mixed Methods}

\author{
Ant\^{o}nio Tadeu A. Gomes,
Weslley S. Pereira,
Fr\'{e}d\'{e}ric Valentin
}
\address{LNCC - Laborat\'orio Nacional de Computa\c c\~ao Cient\'ifica\\
25651-075 Petr\'opolis, RJ, Brasil}
\email{\{atagomes,weslleyp,valentin\}@lncc.br}

\author{
Diego Paredes
}
\address{Instituto de Matematicas\\
PUCV - Pontificia Universidad Catolica de Valparaiso\\
Chile}
\email{diego.paredes@pucv.cl}

\begin{document}

\lstset{language=C++,basicstyle=\tiny\ttfamily,breaklines=true,frame=lines,numbers=left,escapechar=|}

\begin{abstract}
The family of Multiscale Hybrid-Mixed (MHM) finite element methods has received considerable attention from the mathematics and engineering community in the last few years.  
The MHM methods allow solving highly heterogeneous problems on coarse meshes while providing solutions with high-order precision. It embeds independent local problems which are responsible for upscaling unresolved scales into the numerical solution. These local contributions are brought together through a global problem defined on the skeleton of the coarse partition. Since the local problems are completely independent, they can be easily computed in parallel. 
In this paper, we present two simulator prototypes specifically crafted for the MHM methods, which adopt two different implementation strategies: 
\begin{inparaenum}[(i)]
\item a multi-programming language approach, each language tackling different simulation issues; and
\item a classical, single-programming language approach.
\end{inparaenum}
Specifically, we use C++ for numerical computation of the global and local problems in a modular way; for process distribution in the simulator, we adopt the Erlang concurrent language in the first approach, and the MPI standard in the second approach.
The aim of exploring these different approaches is twofold:
\begin{inparaenum}[(i)]
\item allow for the deployment of the simulator both in high-performance computing  (with MPI) and in cloud computing environments (with Erlang); and
\item pave the way for further exploration of quality attributes related to software productivity and fault-tolerance, which are key to Exascale systems.
\end{inparaenum}
We present a performance evaluation of the two simulator prototypes taking into account their efficiency.  
\end{abstract}

\maketitle

\section{Introduction}

The numerical simulation of multiscale phenomena, such as fluid flows in porous media and wave propagation in nano-structures, provides effective ways to solve relevant problems in areas such as energy, climate/weather, and health.
In this context, the use of Finite Element Methods (FEM) is particularly disseminated in industry and academia to simulate these phenomena. 
Despite the fact that classical FEM methods~(e.g.\ the Galerkin method) are appropriate for many different problems, their accuracy may seriously decay when they face problems with highly multiscale features.
Because of that, many modern FEM techniques have emerged~\cite{Hou1997,Efendiev2015}. 
Among them, of particular interest to this paper, is the family of Multiscale Hybrid-Mixed (MHM) methods originally proposed in~\cite{harder2013,AraHarParVal13a}.  
From a mathematical viewpoint, the MHM methods naturally incorporate multiple scales (the MHM-levels) and provide solutions with high-order precision on coarse meshes. 
From a computational viewpoint, at a lower MHM-level a set of completely independent \textit{local} problems can be easily computed in parallel to provide information to its upper MHM-level (the so-called \textit{global} problem). As a result, the MHM methods are naturally shaped to be used in parallel computing environments. Thereby, the MHM methods appear as a highly competitive option to handle realistic multiscale boundary value problems (BVPs).

The MHM methods have been numerically validated for problems such as the Darcy equation~\cite{harder2013}, the Stokes/Brinkman models \cite{AraHarPozVal16}, the advective-reactive dominated problems~\cite{harder2015}, and the linear elasticity problem~\cite{HarMadVal16}.
In this paper, we present and evaluate two simulator prototypes specifically crafted for the MHM methods, which adopt two different implementation approaches: 
\begin{inparaenum}[(i)]
\item a multi-programming language approach with dynamic process scheduling; and
\item a single-programming language approach with static process scheduling.
\end{inparaenum}
Specifically, we use C++ for numerical computation of the global and local problems in a modular way. 
As for process distribution in the simulator, we employ the Erlang concurrent, actor-based language integrated with the C++ modules in the first approach, and a MPI-based C++ application in the second approach.
The aim of exploring these different approaches is twofold:
\begin{inparaenum}[(i)]
\item to allow for the deployment of the simulator both in conventional high-performance computing (HPC) environments (with MPI) and in cloud computing environments (with Erlang); and
\item to pave the way for further exploration of quality attributes related with software productivity and fault-tolerance, which are key to Exascale systems~\cite{CapGeiGropKalKraSin09}.
\end{inparaenum}

We present a preliminary performance evaluation of the two simulator prototypes, taking into account their strong and weak scalability efficiency.
We show that, although not as fast as the MPI version, the Erlang version does provide fairly good speedups up to a few hundred cores, even with all node and link failover mechanisms being set up. 
Besides, despite the lack of formal evidence, it is worth mentioning Erlang's case for software productivity, since it took one of the authors 4 months worth of coding in the Erlang part, from learning a whole new, actor-based programming language to implementing a fully-fledged application with all node and link failover mechanisms enabled.

The remainder of this paper is structured as follows. 
Section~\ref{sec:mhm} overviews the family of MHM methods.
Section~\ref{sec:primitives} presents the key computing primitives of the MHM simulator and how they relate to the algorithm implied by the MHM methods.
Section~\ref{sec:impl} shows the architecture and key design principles that underlie the two different implementations of the simulator.
Section~\ref{sec:eval} offers a performance evaluation of these two simulator implementations. 
Finally, Section~\ref{sec:conc} presents some concluding remarks and perspectives for future work.

\section{The MHM methods}
\label{sec:mhm}

To present the MHM methods in a compact way, we consider the BVP to find $u:\Omega\mapsto\mathbb{R}^d$, $d\in\left\{2,3\right\}$, so that $\mathcal{L}u\,=\,f$ in an open-bounded domain $\Omega$ with  prescribed homogeneous Dirichlet boundary condition on the polygonal boundary $\partial \Omega$. Hereafter  $\mathcal{L}$ is a bounded linear operator.
Some definitions will be needed in the sequel.

\begin{figure}[h!]
   \centering
   \includegraphics[width=0.6\columnwidth]{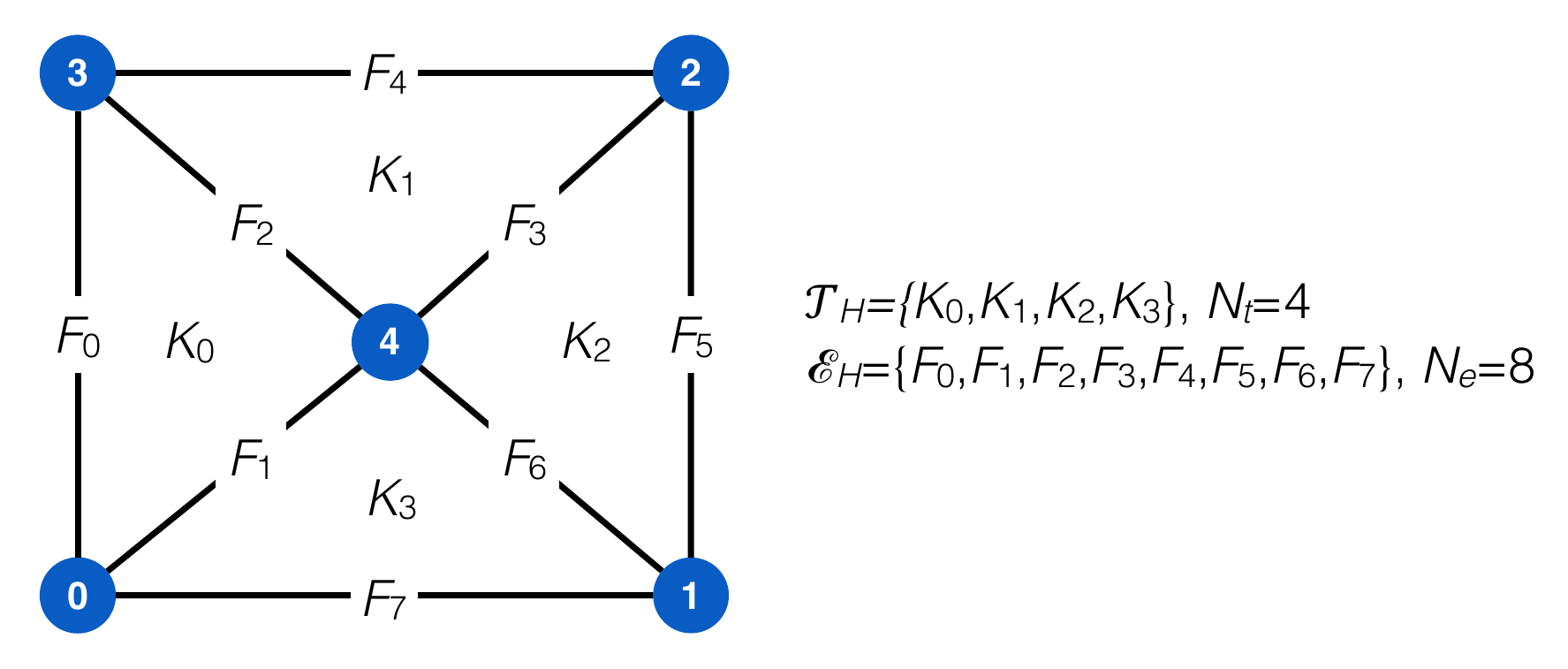} 
   \caption{Mesh example in $\mathbb{R}^2$.}
   \label{fig:mesh}
\end{figure}

First, let $\left\{\trih\right\}_{H\geq 0}$ be a familiy of regular triangulations of $\Omega$ such that each mesh $\trih$ decomposes in a set of globally-indexed elements $\{K_t\}_{t \in T}$, with $T \subset Z^+ = \{0, \dots, N_t-1\}$.
This set defines a \textit{mesh} with $N_t$ elements. 
The label $H$ refers to the \textit{mesh size}, where $H$ is the maximum diameter of any element in the mesh.
Each element $K$ has a boundary $\partial K$ consisting of a set of faces $F$, and some of these faces (those who are internal to the mesh) may be shared between two elements.
Denote by $\partial \trih$ the set of $\partial K$, and by $\fach$ the set in the triangulation of all globally-indexed faces $\{F_e\}_{e \in E}$, with $E \subset Z^+ = \{0,\dots, N_e-1\}$. $\fach$ defines the \textit{skeleton} of a mesh with $N_e$ faces. Figure~\ref{fig:mesh} illustrates these concepts for $\mathbb{R}^2$. We associate each face of $E$ to a unitary normal vector $\mathbf{n}$, taking care to ensure this is facing outward on $\partial \Omega$. Also, the unitary outward normal vector on a face $F$ of element $K$ is represented by $\mathbf{n}^K_F$.

Second, we define some function spaces:
\begin{enumerate} 
\item $V_0$ is the null space (with basis $\{\upsilon_0, \dots, \upsilon_{N_{V_0}-1}\}$) of operator $\mathcal{L}\,|_K$ for all $K \in \trih$ (e.g., for the Darcy equation  $V_0$ is formed by piecewise constants in each $K$~\cite{harder2013}, and then $N_{V_0}=N_t$.);
\item $V_0^\perp
:=\bigoplus_{K \in \trih}V_0^\perp(K)$ is the orthogonal complement of $V_0$ such that it yields the wellposedness of $\mathcal{L}\,|_K$ for all $K\in \trih$;
\item $V
:=\bigoplus_{K \in \trih}V_{K}$ is the space defined as the direct sum between $V_0$ and $V_0^\perp
$  (e.g., for the Darcy equation  $V_{K}=H^1(K)$~\cite{harder2015}); 
\item $\Lambda$ is the space containing the Lagrange multipliers defined over $\partial \trih$~\cite{AraHarParVal13a}.
\end{enumerate} 
For these spaces, we also define the $L^2$ inner products $(\cdot,\cdot)_{\trih}:=\sum_{K \in \trih}(\cdot,\cdot)_K$ and $(\cdot,\cdot)_{\partial \trih}:=\sum_{K \in \trih}(\cdot,\cdot)_{\partial K}$, respectively.

Third,  we define the following finite-dimensional space needed for approximating the exact field $u$: 
\begin{itemize}  
\item[(i)] $\Lambda_H^{m,l} \subset \Lambda$  is the space of discontinuous  polynomial functions on $F_e \in E$ of degree less than or equal to $l\geq 0$ and defined on a uniform partition of $F_e$, composed of $m$ elements ($m\geq 1$), with basis $\{\psi_0, \dots, \psi_{L_l-1}\}$. 
\end{itemize} 

\begin{figure*}[h!]
\centering
\subfigure[Global problem.]{\includegraphics[width=0.45\columnwidth]{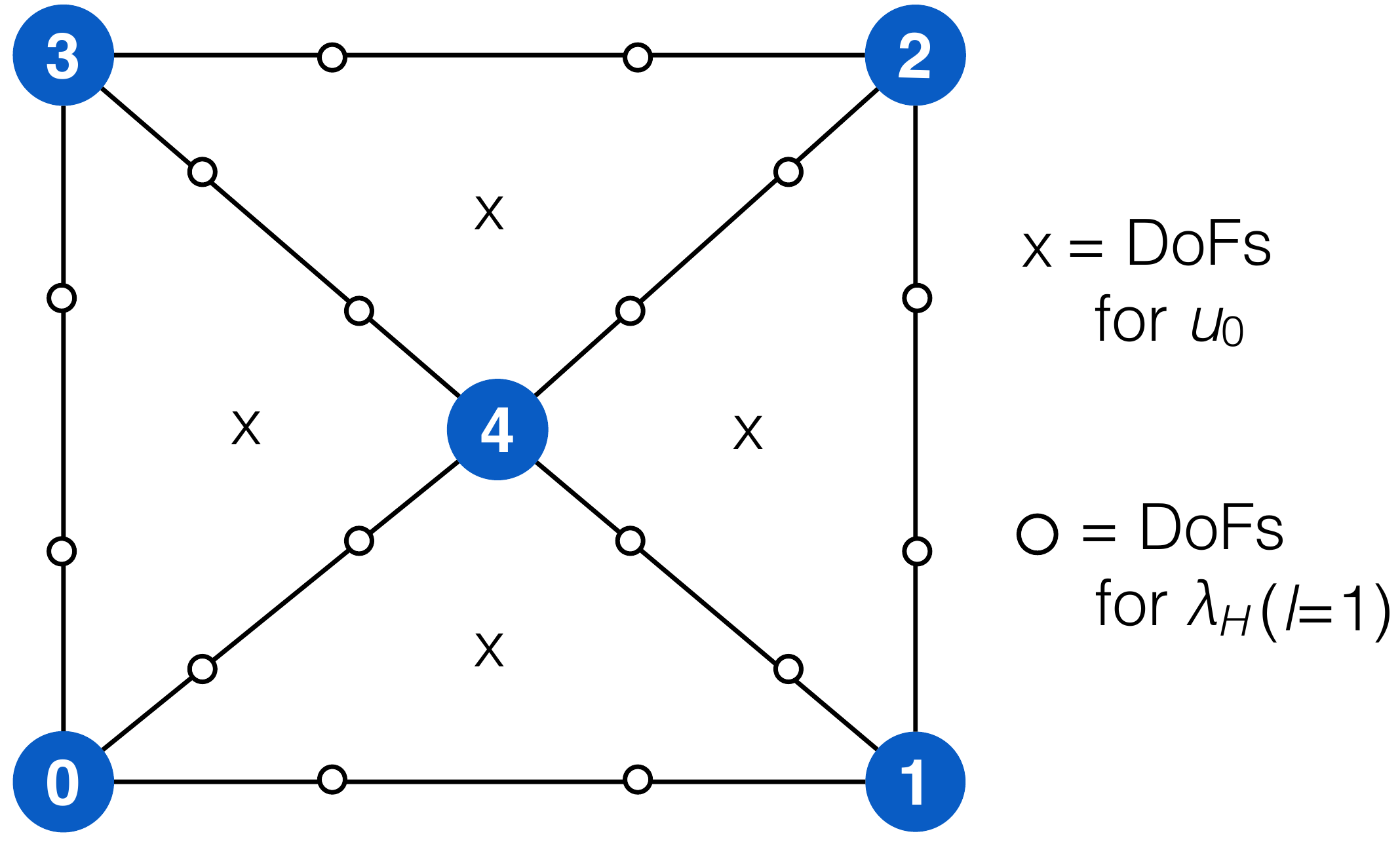}}
\subfigure[Local problem.]{\includegraphics[width=0.5\columnwidth]{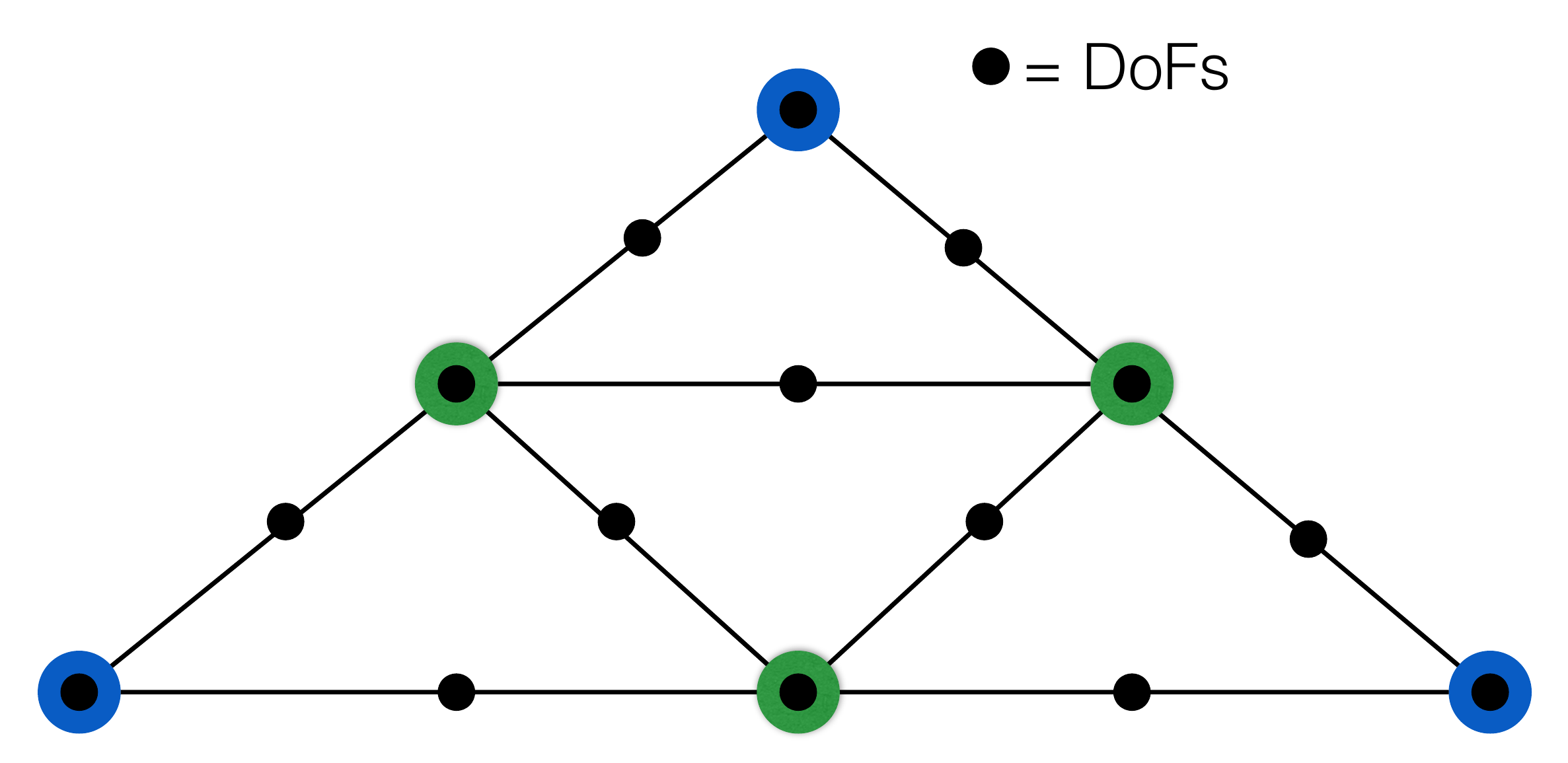}}
\caption{DoFs distribution in MHM.} 
\label{fig:dofs}
\end{figure*}

Given these definitions, a BVP can be written within the MHM terminology in terms of a collection of local problems---each $K$ being associated with a different local problem---brought together by a global problem defined on $\trih$. Specifically, the global formulation is as follows:
Find $(\lambda_H,u_0) \in \Lambda_H^{m,l} \times V_0$ such that
\begin{equation} 
\label{eq:global}
\left\{
\begin{array}{rcl}
(T_h\lambda_H,\mu_l)_{\partial \trih} + (u_0,\mu_l)_{\partial \trih}  & = &  - (\hat{T}_hf,\mu_l)_{\partial \trih}\\ 
(\lambda_H,q_0)_{\partial \trih}  & =  &  (f,q_0)_{\trih} 
\end{array}
\right.
\end{equation} 
\noindent for all $(\mu_l,q_0) \in \Lambda_H^{m,l} \times V_0$. $L_l$ is the total number of degrees of freedom (DoFs) for $\Lambda_H^{m,l}$, and its value depends on the adopted degree $l$ of the piecewise polynomials in $\Lambda_H^{l,m}$, the total number of partitions $m$ in each face $F_e$, and the total number of faces $N_e$ in the mesh.
In the above formulation, $f$ is given; $T_h$ and $\hat{T}_h$ are linear bounded operators driven by the local problems.
Figure~\ref{fig:dofs}(a) illustrates an example of the distribution of the DoFs for $u_0$ in the case of the Darcy equation and $\lambda_H$ (with $l=1$ and $m=1$) in the mesh of Figure~\ref{fig:mesh}.

The local operators $T_h$ and $\hat{T}_h$ in \eqref{eq:global} are responsible for the upscaling process. To precise them, let $V_h(K_t) \subset V_0^\perp (K_t)$ be a finite-dimensional space (with basis $\{\phi_0^t, \dots, \phi_{D_{K_t}-1}^t\}$) defined within an element $K_t$ in the mesh.
At the local level, each element $K_t$ is considered a domain on its own and as such may have an internal triangulation, thus defining a \textit{submesh} within $K_t$. 
Each of these submeshes may have a different characteristic size $h\geq 0$, depending on the high-contrast aspects of the coefficients involved in operator $\mathcal{L}$ for each $K_t$.
The number of DoFs $D_{K_t}$  in $V_h(K_t)$ will therefore depend on the degree of the polynomial functions  in $V_h(K_t)$ and $h$.  
For the purposes of this paper, we assume the solution to each local problem is approximated using the continuous Galerkin method with a Lagrange basis for $\mathbb{P}_h^k$ (the space of piecewise polynomials of degree less than or equal to $k$); $D_{K_t}$ is then determined by the degree $k$ of the polynomials and the number of elements in the submesh. As such, we employ the bilinear form $a(u_h,w_h)_{K_t}$ to set up operators $T_h$ and $\hat{T}_h$. It is defined (as usual) as the action of $\mathcal{L}u_h$ on $w_h$ in $K_t$, with $u_h,w_h \in V_h(K_t)$.  Figure~\ref{fig:dofs}(b) illustrates an example of $V_h(K_t)$ with quadratic interpolation $\mathbb{P}_h^{2}$ for element $K_3$ in the mesh of Figure~\ref{fig:mesh}.
In the example, $K_3$ has been internally triangulated generating a submesh with 4 elements.

Each local problem comprises a set of local \textit{subproblems}.
One such subproblem defines $\hat{T}_h$ by computing  $\eta_t^f :=  \hat{T}_h f\in V_h(K_t)$  in each $K_t\in\trih$ as follows:
Find $\eta_t^f \in V_h(K_t)$ such that 
\begin{equation}
\label{eq:etaf}
a(\eta_t^f,w_h)_{K_t}  = (f,w_h)_{K_t}\quad\text{for all }w_h\in V_h(K_t). 
\end{equation}

The other $N_I$ subproblems define $T_h$ by computing  $\eta_t^i:=T_h \psi_i$ from the following local problems: For $i \in \{0, \dots, N_I-1\}$, find $\eta_t^i \in V_h(K_t)$ such that 
\begin{equation}
\label{eq:etai}
a(\eta_t^i,w_h)_{K_t}  = -(\psi_i,w_h)_{\partial K_t}\quad\text{for all } w_h \in V_h(K_t), 
\end{equation}
where $\psi_i$ changes its sign accordingly to $\mathbf{n}\cdot\mathbf{n}^{K_t}_F$, and $N_I$ is defined by the number of faces in $K_t$ and by the degree $l$ of the polynomials and $m$ in $\Lambda_H^{m,l}$.
Finally, from the solution of Equations~\eqref{eq:global}, \eqref{eq:etaf} and \eqref{eq:etai}, and using that $\lambda_H = \sum_{i=0}^{N_I-1} k_i  \psi_i$ with $k_i\in\mathbb{R}$, the exact field $u$ is approximated by
\begin{align}
\label{eq:u_Hh}
u_{H,h} &:= u_0 + T_h \lambda_H + \hat{T}_h f \nonumber \\
& = u_0 + \sum_{K_t\in\trih} \Big[\sum_{i=0}^{N_I-1} k_i \eta^i_t + \eta_t^f\Big].
\end{align}
\section{The simulator primitives}
\label{sec:primitives}

The MHM simulator is built on top of key computing primitives that bear correspondence with the major mathematical building blocks of the MHM method: \textsc{splitProblem}, \textsc{solveLocalProblem}, \textsc{reduceLocalProblems}, and \textsc{computeSolution}.
These primitives are combined to implement the MHM simulator as depicted abstractly in Algorithm~\ref{alg:main}.
\begin{algorithm*}
\caption{Pseudocode of \textit{main} procedure for the MHM simulator.}
\label{alg:main}
\begin{algorithmic}
\Require $\trih = \{K_t\}_{t \in T}$ \Comment{Mesh with indexed elements}
\State $P_{\trih} \gets$ \Call{splitProblem}{$\trih$} \Comment{$P_{\trih}:=\{(K_t, \{F_t\}\}_{t \in T}$}
\State $S_{\trih} \gets \emptyset$
\ForAll{$p_t \in P_{\trih}$} \Comment{$p_t:=(K_t, \{F_t\})$}
	\State $s_t \gets$ \Call{solveLocalProblem}{$p_t$} \Comment{$s_t:=(\eta^f_t, \{\eta^i_t\}_{i \in \{0, \dots, N_I-1\}})$}
	\State $S_{\trih} \gets S_{\trih} \cup s_t$
\EndFor
\State ($u_0,\lambda_H) \gets$ \Call{reduceLocalProblems}{$S_{\trih}$} 
\State $u_{H,h} \gets $ \Call{computeSolution}{$u_0$, $\lambda_H$, $S_{\trih}$}
\end{algorithmic}
\end{algorithm*}

The \textsc{splitProblem} primitive divides the original problem posed over a mesh resulting from a triangulation $\trih$ into a global problem (the 1\textsuperscript{st} MHM-level) and a set of local problems (the 2\textsuperscript{nd} MHM-level), one for each element $K_t$ in the mesh.
This primitive returns a set $P_{\trih}$ of pairs $(K_t, \{F_t\})$, where $\{F_t\}$ represents the set of all faces of $K_t$.

The \textsc{solveLocalProblem} computes the contributions $\eta_t^f$ and $\{\eta_t^i\}_{i \in \{0, \dots, N_I-1\}}$ from a specific element $K_t$ in the mesh to be used in the global problem. 
Functions $\eta_t^f$ and $\eta_t^i$, $i \in \{0, \dots, N_I-1\}$ are searched as the solution of \eqref{eq:etaf} and \eqref{eq:etai}, respectively, in the following form:
\begin{gather}
\label{eq:etaf_comb}
\eta_t^f=\sum_{j=0}^{D_{K_t}-1} c_j^t \phi_j^t\quad\text{and}\quad\eta_t^i=\sum_{j=0}^{D_{K_t}-1} d_{i,j}^t \phi_j^t .
\end{gather}
%

The coefficients $c_j^t$ and $\{d_{i,j}^t\}_{i \in \{0, \dots, N_I-1\}}$ are the actual numerical results computed by the \textsc{solveLocalProblem} primitive through the solution of a set of linear systems of the form $\matr{A}\mathbf{C}=\mathbf{F}$ and $\{\mathbf{\underline{A}}\mathbf{D}_i=\mathbf{N}_i\}_{i \in \{0, \dots, N_I-1\}}$.
It is worth noting that these systems share the same matrix $\mathbf{\underline{A}} := [a(\phi_j^t,\phi_i^t)]_{i,j \in \left\{0, \dots, D_{K_t}-1\right\}}$, albeit with different load vectors.
This feature means that $\mathbf{\underline{A}}$ needs to be factorized only once (for linear problems), and therefore the increase in the value of degree $l$ or $m$ in the approximation space $\Lambda_H^{m,l}$ only increases the number of matrix-vector multiplications, which can be trivially parallelized even within the context of a single local problem by means of vectorized computations.
Moreover, if the coefficients associated with the operator $\mathcal{L}$ are not time-dependent, $\mathbf{\underline{A}}$ needs to be assembled and factorized only during the first timestep in a time marching scheme. 

The \textsc{reduceLocalProblems} primitive gathers the contributions $\eta_t^f$ and $\{\eta_t^i\}_{i \in \{0, \dots, N_I-1\}}$ from the subproblems to compute $u_0$ and $\lambda_H$ from Equation \eqref{eq:global}. The underlying  linear system associated to \eqref{eq:global} reads:
\begin{equation}
\label{eq:mat_global}
\left( \begin{array}{ccc}
\mathbf{\underline{A}} & \mathbf{\underline{B}}^{\mathrm{T}} \\
\mathbf{\underline{B}} & \mathbf{\underline{0}} \end{array} \right) 
\left( \begin{array}{ccc}
\mathbf{L}\\
\mathbf{P} \end{array} \right) 
=
\left( \begin{array}{ccc}
\mathbf{E} \\
\mathbf{F} \end{array} \right), 
\end{equation}
\noindent with:
\begin{equation*}
\begin{array}{l}
\mathbf{\underline{A}} := [(\psi_j, \eta^i_t)]_{i,j \in \left\{0,\dots,L_l-1\right\}},\\
\mathbf{\underline{B}} := [(\psi_j, \upsilon_i)]_{i \in \left\{0,\dots,N_{V_0}-1\right\}, j \in \left\{0,\dots,L_l-1\right\}},\\
\mathbf{E} := [-(\psi_j, \eta_t^f)]_{j \in \left\{0,\dots,L_l-1\right\}},\\
\mathbf{F} := [(\upsilon_j, f)]_{j \in \left\{0,\dots,N_{V_0}-1\right\}}.
\end{array}
\end{equation*}

The computed entries $[k_j]_{j \in \{0,\dots,L_l-1\}}$ in $\mathbf{L}$ define the solution to $\lambda_H$ in $\Lambda_H^{m,l}$; likewise, the computed entries $[b_j]_{j \in \{0,\dots,N_{V_0}-1\}}$ in $\mathbf{P}$ define the solution $u_0$ in $V_0$ through
\begin{gather}
\label{u00}
u_0 = \sum_{j=0}^{N_{V_0}-1} b_j \upsilon_j.
\end{gather}

Finally, owing to the \textsc{computeSolution} primitive, the two-level numerical solution $u_{H,h}$ reads
\begin{equation*}
u_{H,h} = \sum_{j=0}^{N_{V_0}-1} b_j \upsilon_j + \sum_{K_t\in\trih} \sum_{j=1}^{D_{K_t}-1} \Big[\sum_{i=0}^{N_I-1} k_i d_{i,j}^t \phi_j^t +   c_j^t\phi_j^t \Big].
\end{equation*}
where we used  \eqref{u00} and \eqref{eq:etaf_comb} in \eqref{eq:u_Hh}.
\section{Implementation details}
\label{sec:impl}

\begin{figure*}[h!]
   \centering
   \subfigure[Erlang version.]{\includegraphics[width=0.98\columnwidth]{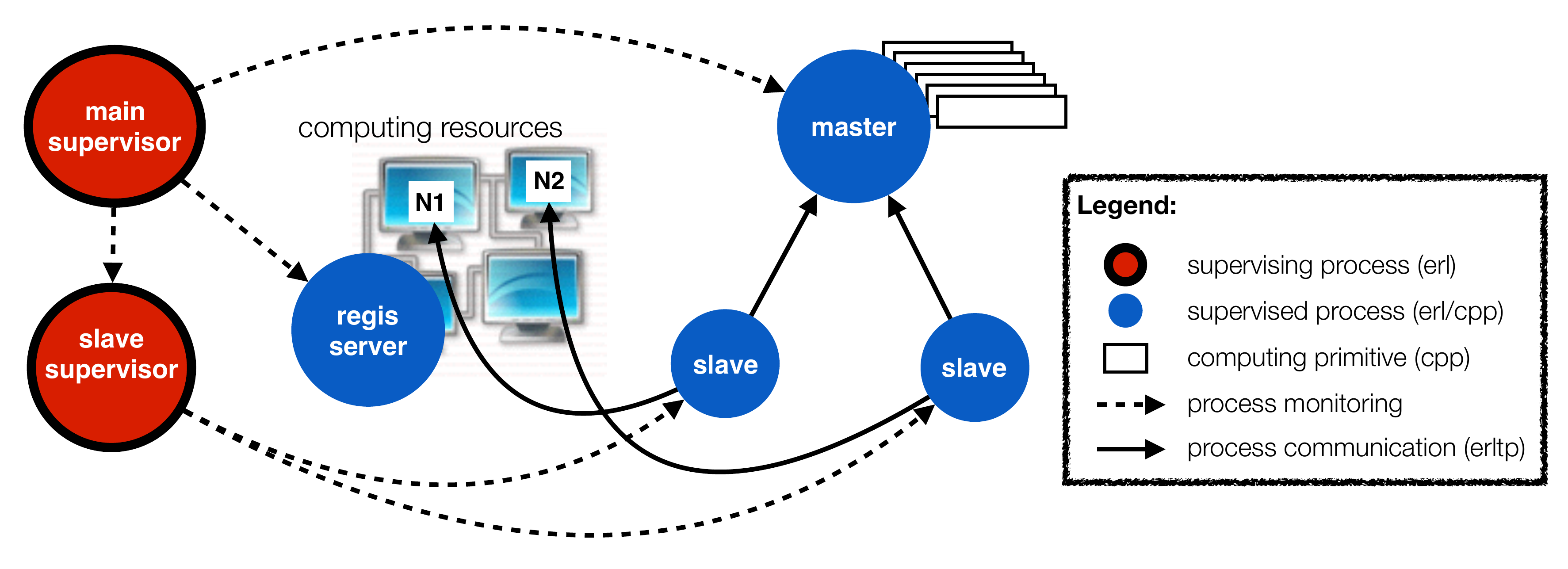}}
   \subfigure[MPI version.]{\includegraphics[width=0.85\columnwidth]{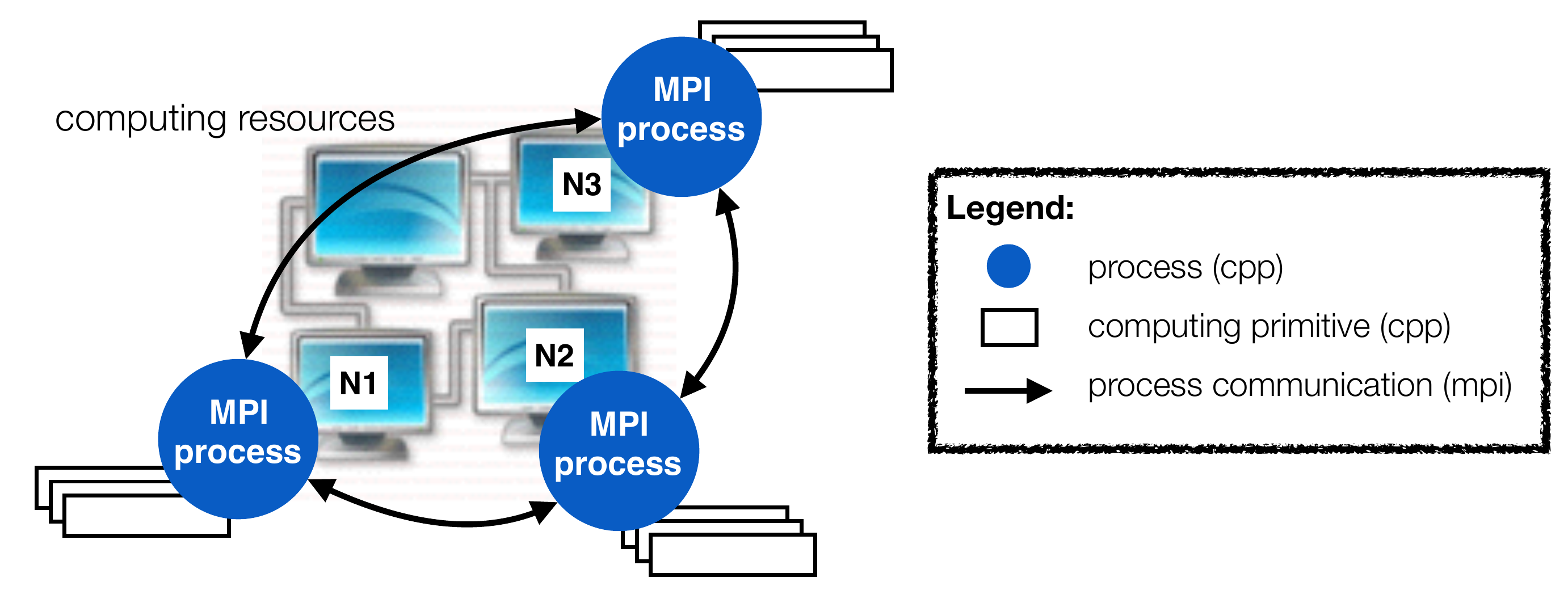}}
   \caption{Architecture of the MHM simulator.}
   \label{fig:architecture}
\end{figure*}

The MHM simulator is a distributed system.
It exploits either Erlang's or MPI support for scalable computing
and C++'s efficiency for the numerical computations of the global and local problems.
The overall architecture of the MHM simulator is depicted in Figures~\ref{fig:architecture}(a) and \ref{fig:architecture}(b) for the Erlang and MPI versions, respectively.
The following subsections describe the two ``layers'' (distribution and computation) of the architecture.

\subsection{The distribution layer}

\subsubsection{The Erlang version}

Before describing this version of the distribution layer, we give a short introduction to the Erlang language and its runtime, including its fault-tolerance mechanisms.

\textbf{An Erlang prime.} Erlang is a concurrent language with functional programming features.
It is a language that compiles to an intermediate format (\textit{byte codes}) and runs in virtual machines (the Erlang \textit{nodes}).
Typically, each Erlang node runs on a different physical machine.

Erlang programs are built on functions that can be called from the same process of the calling function or dispatched (\textit{spawned}) to another process residing in the same or a different Erlang node.
Processes in Erlang run concurrently and communicate with each other by message passing.
They usually follow a common pattern: once spawned they call a tail-recursive function to process and produce data.
This data is referred to as the process \textit{state}.
The tail-recursive function usually takes the following sequence of operations: 
\begin{inparaenum}[(i)]
\item receive a message from another process, 
\item handle the message, 
\item update the process state, and 
\item pass the updated state to a new invocation of the function through its tail-recursive call.
\end{inparaenum}
Finally, a special \textit{stop} message is usually defined to allow the receiving process to be terminated gracefully.

The Erlang runtime provides templates---called \textit{behaviors}---that implement the above pattern for processes acting as servers (the \texttt{gen\_server} behavior) and finite state machines (FSMs -- the \texttt{gen\_fsm} behavior), among others.
Building on its messaging mechanism, Erlang offers process monitoring capabilities that allow a process to detect whether another process has failed and act accordingly. 
This is explored in an specific behavior called \texttt{supervisor}.
A supervisor is a process whose main task is to manage---spawn, monitor, and eventually re-spawn---other processes in case of failure.
Supervisors may manage any type of Erlang process, including other supervisors.
This organization creates what is called a \textit{supervision tree}.
Finally, an Erlang \textit{application} is a special kind of behavior that allows an entire supervision tree to be started and stopped as a unit.
To provide tolerance to hardware faults, an Erlang application can be controlled in a distributed manner, by employing two or more Erlang nodes running in different machines.
This setup configures a \textit{distributed application}.
If the Erlang node where a distributed application goes down (e.g. because of a network partition or machine failure), the application can be restarted at another node.

\noindent\textbf{The simulator distributed application}. Figure~\ref{fig:architecture}(a) illustrates the chief processes of the Erlang-based simulator distributed application.
The \textit{main supervisor} implements the \texttt{supervisor} behavior.
It manages a \textit{master}, a \textit{regis server}, and a \textit{slave supervisor}, which are all spawned at the application bootstrap.
All these processes run in a single Erlang node; if this node becomes unavailable, they are resumed in another Erlang node with their state preserved thanks to the use of the Mnesia fault-tolerant database embedded in the Erlang runtime.
This is the first key feature that allows the Erlang-based simulator to have a very fine-grained fault tolerance support.

The master implements the \texttt{gen\_server} behavior.
It keeps track of a set of \textit{computing tasks}, which are dynamically scheduled to the available computing resources, as described below.
Each task is related to either a global or a local problem in the MHM methods, and has an associated state---``not solved'', ``being solved'', ``being reduced'' (for global problems only) or ``solved''.
Besides, each task maps onto an instance of the simulator primitives described in Section~\ref{sec:primitives}.
These primitives, which are implemented in C++ (see Section~\ref{subsec:cpp}), run on computing resources managed by the regis server, which also implements the \texttt{gen\_server} behavior.
The regis server collects status information about the computing resources in the execution platform (e.g.\ computing nodes in an HPC cluster or virtual machines in a cloud) and, for each such resource, asks the slave supervisor (another implementation of the \texttt{supervisor} behavior) to dynamically create/destroy new \textit{slaves} associated with the resource whenever the resource becomes available/unavailable.
This is the second key feature for the fine-grained fault tolerance support in the Erlang-based simulator.

The slaves implement the \texttt{gen\_fsm} behavior.
They run in the same Erlang node as the other processes of the Erlang application.
Once started, an slave keeps on asking the master about new computing tasks.
The master replies to a request from a slave by:
\begin{inparaenum}[(i)]
\item selecting an appropriate task, 
\item changing the state of the task to ``being solved''/``being reduced'', and 
\item replying to the slave with the simulator primitive to be executed.
\end{inparaenum}
The slave then dispatches the execution of the primitive onto its associated computing resource, and keeps track of the primitive execution.
When the slave detects that the primitive is finished, it sends the results of the primitive to the master, which may change the state of the task or create new tasks.
If the slave detects that the primitive hasn't completed because of a failure in a computing resource, it notifies the master and the regis server, which dynamically reallocate the task to a different resource.

\begin{figure}[h!]
   \centering
   \includegraphics[width=\columnwidth]{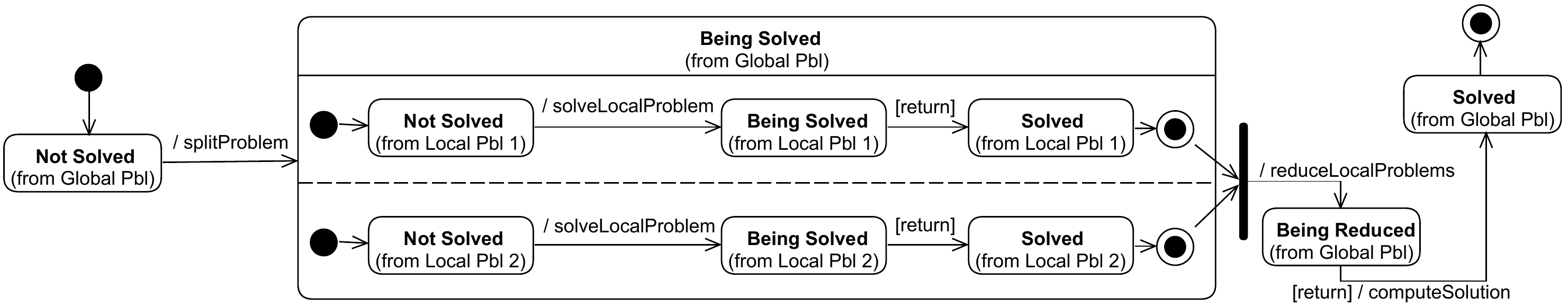} 
   \caption{Evolution of task states.}
   \label{fig:task_states}
\end{figure}

Figure~\ref{fig:task_states} illustrates how the state of the tasks evolves in the simulator.
The simulator starts with a single task representing a global problem---$\trih$ in Algorithm~\ref{alg:main}---being published in the master with the ``not solved'' state.
At this point a single slave will be able to get this task, which triggers the dispatching of the \textsc{splitProblem} primitive to its associated computing resource and the changing of its state to ``being solved''. 
The result of this primitive---$P_{\trih}$ in Algorithm~\ref{alg:main}---is taken by the slave as a set of requests---one for each $p_t \in P_{\trih}$---for the creation of new tasks in the master.
Each of these new tasks is related to a different local problem and added to the master with the ``not solved'' state.
To simplify the illustration, in Figure~\ref{fig:task_states} we only present two local problems being derived from the \textsc{splitProblem} primitive.

From this point on the slaves start to get different tasks related with local problems, changing their states to ``being solved'' and dispatching the execution of the \textsc{solveLocalProblem} primitive to their associated computing resources.
We employed the sequential LU factorization implementation of the Eigen library\footnote{\url{http://eigen.tuxfamily.org}} within this primitive for solving the linear systems that compute $c_j^t$ and $\{d_{i,j}^t\}_{i \in \{0, \dots, N_I-1\}}$ in Equation~\eqref{eq:etaf_comb}, with the intention to make as many local problems be solved in parallel in a single computing resource as the amount of cores the resource provides.
When these primitive invocations are finished, the slaves send their results---$s_t \in S_{\trih}$ in Algorithm~\ref{alg:main}---to the master, which changes the state of their corresponding tasks to ``solved''.
The slaves then proceed with getting other tasks and dispatching them to their resources until there are no more tasks in the ``not solved'' state.

Once all local problems are in the ``solved'' state, the next request from an slave is replied with the \textsc{reduceLocalProblems} primitive, and the state associated with the global problem is changed from ``being solved'' to ``being reduced''.
Again, we employed the LU factorization within this primitive to solve the linear system in Equation~\eqref{eq:mat_global}.
For the \textsc{reduceLocalProblems} primitive, however, we used the shared-memory parallel Pardiso solver.\footnote{\url{http://www.pardiso-project.org}}
The result from the \textsc{reduceLocalProblems} primitive triggers the immediate execution of the \textsc{computeSolution} primitive and a changing of the task's state to ``solved'' in the master.
After that, the simulator finishes.

\subsubsection{The MPI version}

Figure~\ref{fig:architecture}(b) illustrates the chief processes of the MPI-based application.
The $R$ MPI processes that make part of the application are omnipotent; in contrast with the Erlang version, there's no master or slave process. 
Each MPI process $r \in R$ is responsible for running its own \textit{share} of the \textsc{splitProblem} primitive. 
This share corresponds to the execution of \textsc{splitProblem} over a partition $\trihr$ of $\trih$. 
Such a partitioning configures a static process scheduling that is kept until the end of the simulation.

The result of the \textsc{splitProblem} primitive at each MPI process ($P_{\trihr}$) is taken by this process as a set of local problems to be solved---one for each $p_t \in P_{\trihr}$.
From this point on the MPI processes start to execute the \textsc{solveLocalProblem} primitive for all $p_t$ they are responsible for, in the same manner as the Erlang version.

An MPI barrier is used for making all MPI processes wait for each other as they finish solving all their corresponding local problems.
Once all local problems are solved, each MPI process $r \in R$ is responsible for running its own share of the \textsc{reduceLocalProblems} primitive.
This is where the MPI version of the MHM simulator mostly differs from the Erlang counterpart.
A share of the \textsc{reduceLocalProblems} primitive corresponds to the distributed assembly of the global linear system shown in Equation~\eqref{eq:mat_global}, and its solution through LDL$^T$ factorization using the distributed interface of the parallel PaStiX library~\cite{HRR01a}.
The result from the \textsc{reduceLocalProblems} primitive triggers at each MPI process the immediate execution of its own share of the \textsc{computeSolution} primitive.
After that, the simulator finishes.

As it can be seen from Figure~\ref{fig:architecture}(b), this implementation is structurally simpler than the one of the Erlang-based simulator.
Most of such simplicity is due to the lack, in this implementation, of a dynamic process scheduling scheme and of any finer-grained fault-tolerance mechanisms.
This means that a failure in a single computing resource running an MPI process crashes the whole application.
In addition, checkpoints for the fast restart of a crashed application are only available in two points: at the end of the \textsc{splitProblem} primitive, and at the end of all invocations of the \textsc{solveLocalProblem} primitive.
Therefore, a failure during the solution of a single local problem results in \textit{all} local problems being solved again, but if all local problems have been solved, the application can be restarted from the point of the  \textsc{reduceLocalProblems} primitive.

\subsection{The computation layer}
\label{subsec:cpp}

C++ is used for implementing the numerical computations that ultimately solve the global and local problems. 
The C++ layer is divided into 2 packages. 

\begin{figure}[h!]
   \centering
   \includegraphics[width=0.7\columnwidth]{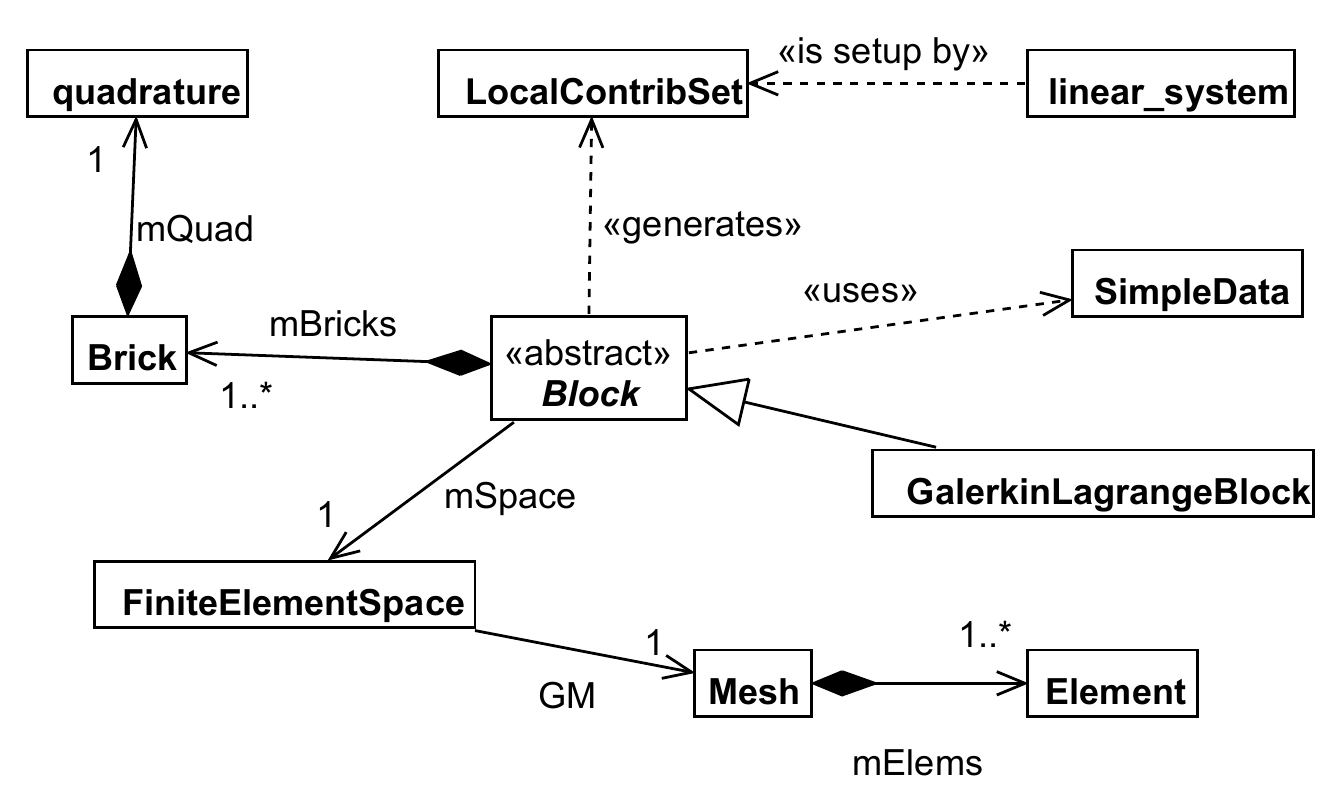} 
   \caption{UML diagram of main C++ classes.}
   \label{fig:uml_basic}
\end{figure}

The first package implements key concepts present in any FEM-based simulator as C++ classes:
\begin{inparaenum}[(i)]
\item geometric elements (\texttt{Element}) and meshes (\texttt{Mesh});
\item quadrature rules for numerical integration (\texttt{Quadrature});
\item spaces of approximate functions (\texttt{Space});
\item basis function definitions for elements (\texttt{Brick}) and meshes (\texttt{Block}); 
\item coefficient and source term definitions (\texttt{SimpleData});
\item contribution computations (\texttt{LocalContribSet}); and
\item a wrapper to linear solver libraries (\texttt{LinearSolver}).
\end{inparaenum}
Figure~\ref{fig:uml_basic} shows a UML diagram with the main relationships between these classes.

The \texttt{Block} class is an abstract class;
it must be specialized in classes
depending on where the data comes from.
The \texttt{GalerkinLagrangeBlock} class shown in Figure~\ref{fig:uml_basic}, for instance, compute its values from the definitions of basis functions in \texttt{FiniteElementSpace}.

\begin{figure}[h!]
   \centering
   \includegraphics[width=\columnwidth]{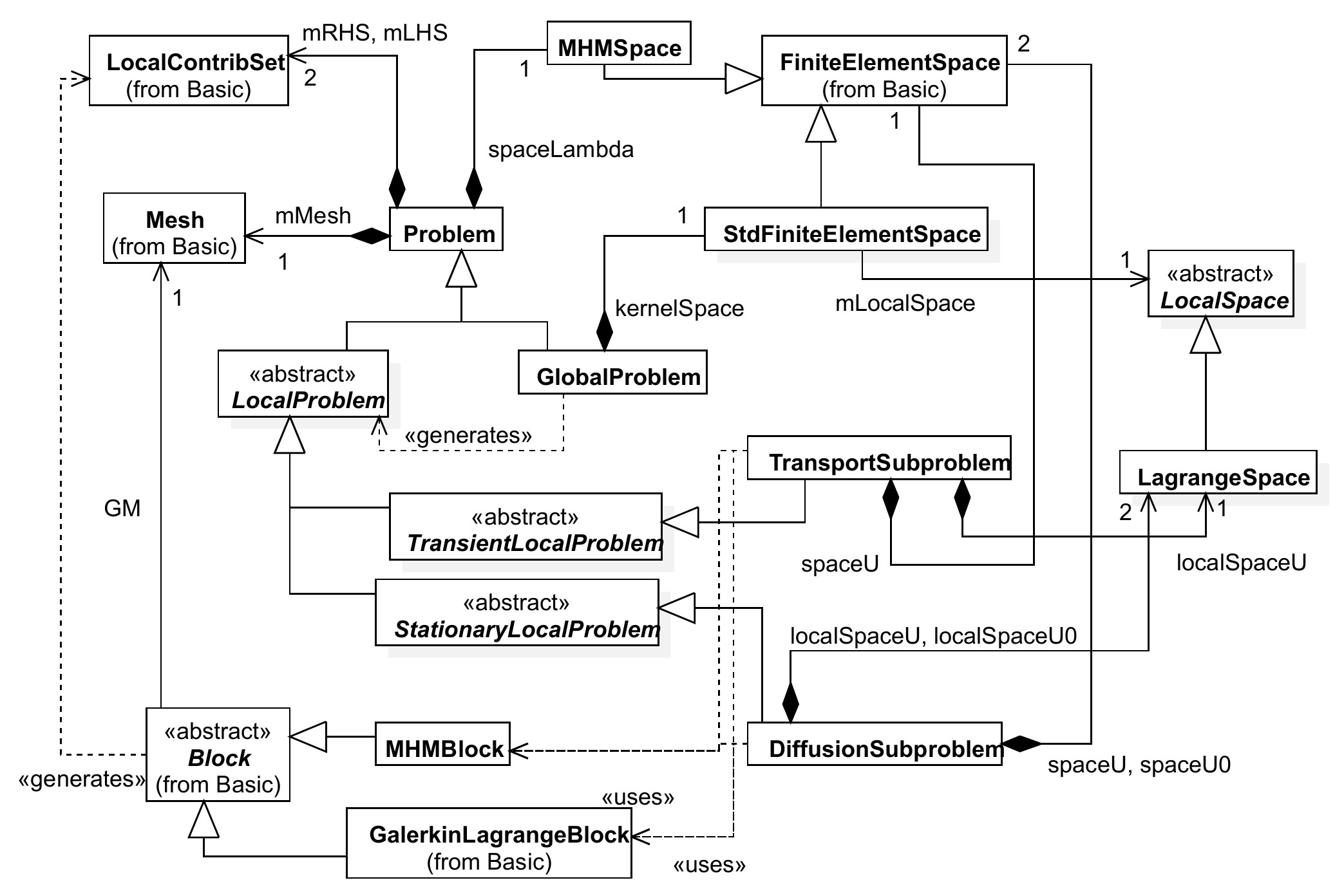} 
   \caption{UML diagram of the C++ classes that implement the framework for MHM methods.}
   \label{fig:uml_mhm}
\end{figure}

The second package comprises a set of C++ classes that implement the framework for MHM methods:
\begin{inparaenum}[(i)]
\item global problem definitions (\texttt{Problem});
\item local problem definitions (\texttt{LocalProblem} and its direct subclasses \texttt{StationaryLocalProblem} and \texttt{TransientLocalProblem});
\item definitions of spaces $V_0$ and $\Lambda_H^{m,l}$ (\texttt{MHMSpace} and subclasses of \texttt{LocalSpace}); and
\item a subclass of Block (\texttt{MHMBlock}).
\end{inparaenum}
Figure~\ref{fig:uml_mhm} presents a UML diagram that depicts the main relationships between these classes.

The \texttt{GlobalProblem} class implements the simulator for Equation~\eqref{eq:global}.
Its methods \texttt{GlobalProblem::split()}, \texttt{GlobalProblem::reduce()}, and \texttt{GlobalProblem::solve()} correspond to the primitives \textsc{splitProblem}, \textsc{reduceLocalProblems}, and \textsc{computeSolution} in Algorithm~\ref{alg:main}.

\texttt{LocalProblem} is an abstract class responsible for:
\begin{inparaenum}[(i)]
\item defining a submesh within an element $K_t\in\trih$,
\item solving a local problem in this submesh, and 
\item computing the local contributions related with this submesh to the global problem. 
\end{inparaenum}
Its main method---\texttt{LocalProblem::solveLocalProblem()}---corresponds to primitive \textsc{solveLocalProblem} in Algorithm~\ref{alg:main}.
The subclasses \texttt{StationaryLocalProblem} and \texttt{TransientLocalProblem} specialize the behavior of this method to  stationary and transient problems.
Specializations of these two subclasses then override \texttt{solveLocalProblem()} to provide concrete behavior.
For instance, the \texttt{DiffusionSubproblem} class specializes \texttt{StationaryLocalProblem} to solve the following local subproblems, which particularize the general Equations~\eqref{eq:etaf} and \eqref{eq:etai} for the Darcy equation~\cite{harder2013}: Find $\eta_t^f \in V_h(K_t)$ such that
\begin{equation} 
\label{eq:source_term}
(\kappa\nabla\eta_t^f,\nabla w_h)_{K_t} = (f,w_h)_{K_t}\quad
\text{for all }w_h \in V_h(K_t),
\end{equation}
and for $i \in \{0, \dots, N_I-1\}$, find $\eta_t^i \in V_h(K_t)$ such that
\begin{equation} 
(\kappa\nabla\eta_t^i,\nabla w_h)_{K_t} = -(\psi_i, w_h)_{\partial K_t}\quad
\text{for all }w_h \in V_h(K_t),
\end{equation} 
where $\kappa$ is the diffusion coefficient. 
\section{Performance evaluation}
\label{sec:eval}

We collected strong and weak scaling performance measurements for the two simulator prototypes.
For these measurements, we ran the two prototypes in the Santos Dumont HPC cluster at LNCC in Brazil,  which has the following per-node configuration: 2x CPU Intel Xeon E5-2695v2 Ivy Bridge (12 cores each CPU), 2.4GHZ, 64GB DDR3 RAM.  The nodes of the cluster are interconnected through an Infiniband FDR network with a fat-tree topology, and they share a distributed file system based on Lustre v2.1. 
The C++ code was compiled with Intel's icpc compiler, version 16.0.2 build 20160204, with the optimization flags `$-O2$' and `$-ipo$'  enabled.
We used Eigen version 3.3-rc1, PaStiX version 5.2.2.22, and the Pardiso implementation provided with the MKL library that comes with Intel's Parallel Studio XE 2016.2.062.
The Erlang code was run with Ericsson's Open Source Erlang/OTP system, version 17 erts-6.1, with HiPE (High Performance Erlang) native code compilation enabled.
The MPI implementation used was the one that also comes with Intel's Parallel Studio XE 2016.2.062.

\begin{figure}[h!]
\centering
\subfigure[Mesh with 192 elements including the submeshes.]{\includegraphics[width=0.4\columnwidth]{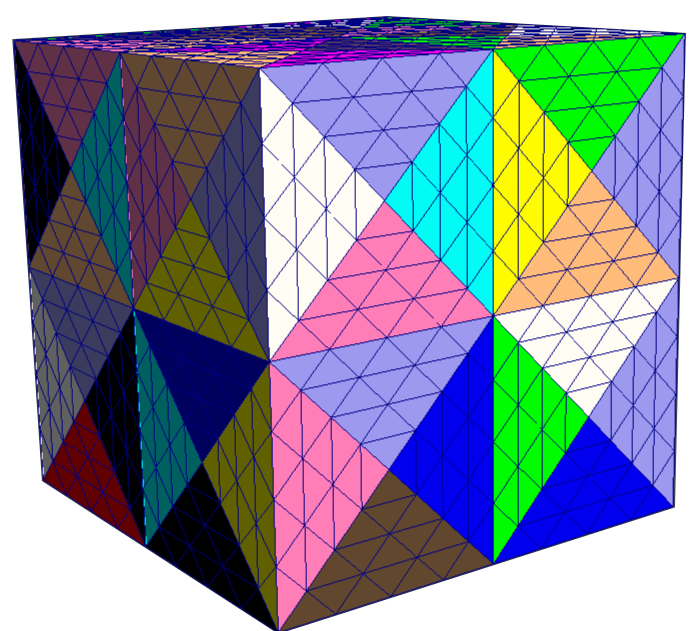}}
\hspace{40pt}
\subfigure[Isolines of the numerical solution at $x=0.75$.]{\includegraphics[width=0.42\columnwidth]{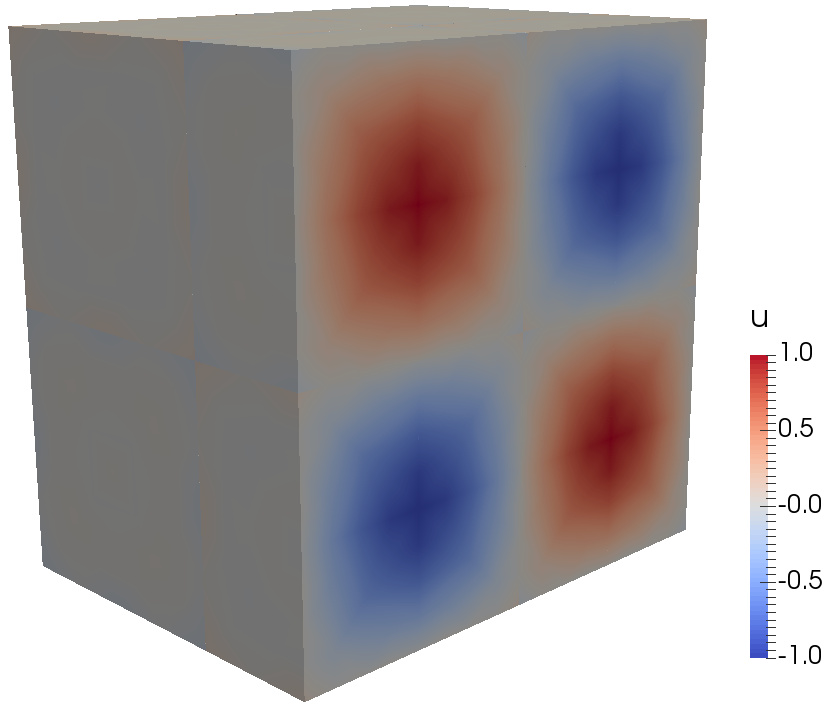}}
\caption{A model problem.} 
\label{fig:simulation3D}
\end{figure}

We solve the Darcy equation defined in the unit cube as a model problem. The diffusion coefficient is set as $\kappa=1$ and  $f(x,y,z) = 12\pi^2\sin(2\pi x)\sin(2\pi y)\sin(2\pi z)$. As such, the exact solution reads $u(x,y,z) = \sin(2\pi x)\sin(2\pi y)\sin(2\pi z)$.
We setup the MHM simulators with 24, 192 and 1,536 elements in a global tetrahedral mesh and within each such element a submesh of either 59 or 6,046 tetrahedra dynamically generated by the TetGen library~\cite{hang2015} during runtime, as part of the computation within the \textsc{solveLocalProblem} primitive. 
We adopted degree $l=0$ and $m=4$ (59 tetrahedra case) or  $m=16$ (6,046 tetrahedra case) for the polynomials in $\Lambda_H^{m,l}$, and degree $k=2$ for the polynomials in the submesh.
We depict the 192 tetrahedra mesh along with its submeshes (59 tetrahedra per element) in Figure~\ref{fig:simulation3D}(a), and the corresponding numerical solution to our model problem in Figure~\ref{fig:simulation3D}(b).


For the sake of illustrative comparison, we present in Table~\ref{tab:galerkin_corr} the number of DoFs that  the standard finite element (Galerkin) method should have to reach a similar approximation order (in terms of accuracy) compared to the one using the MHM methods.
As shown in~\cite{harder2013}, since we are using $l=0$ and $k=2$ with the MHM methods,  the numbers of DoFs in the table correspond   to the use of the linear space $\mathbb{P}_h^{1}$ with the Galerkin method. 
We emphasize that, different from what happens within the MHM methodology, the standard approximation using Galerkin needs to compute all such DoFs at once. This is achieved by solving a single linear system which is typically much larger  than those generated by the global and local problems of the MHM methods.

\begin{table}[b!]
\caption{Correspondence between MHM parameterization and Galerkin DoFs for $\mathbb{P}_h^{1}$.}
\centering
\footnotesize
\begin{tabular}{ll|ll}
& & \multicolumn{2}{c}{Global mesh size} \\ 
& & $1,536$ & $12,288$ \\
\hline
\multirow{2}{1.8cm}{Submesh size} & $59$ & $16,689$ & $127,073$ \\
& $6,046$ & $1,493,697$ & $11,850,113$ \\
\end{tabular}
\label{tab:galerkin_corr}
\end{table}

The setup described above renders 6 different MHM parameterizations, but we do not present any results associated to the configuration with $1,536 \times 59$ tetrahedra.
The reason is that they are much smaller problems and therefore much more subject to statistical variations due to the execution environment.
We tested the other 5 configurations in 3 resource settings: $24$, $192$ and $768$ cores.
For each of the 15 configurations, 3 runs were conducted.
We selected the run with the best overall wall-clock time for presentation herein. 
(It is worth  noticing that no significant statistic variations were identified in the results.)
In the case of the resource setting with $24$ cores, since it occupies a single node in the cluster used for the experiments, we used a baseline version of the MHM simulator that only uses multithreading, without any Erlang or MPI overhead.
The purpose is to  better capture the impact of distribution in the Erlang and MPI versions of the simulator.

Table~\ref{tab:measurements} consolidates all the scalability measurements.
Analyzing the strong scalability, we can observe in the table that the efficiency of both simulators improves as the problems get bigger.
We certainly need  to make  measurements with larger scenarios, but we believe this is a promising result.
Another observation is that the Erlang version of the simulator is less efficient than the MPI one---which was expected---, but not by a large margin.
This is another interesting result because it opens opportunities for  the efficient use of the MHM approach in cloud computing environments, and more generally, for the use of Erlang as an appropriate coordination language for fault-tolerant HPC applications.

\begin{table*}[h!]
\caption{Scalability measurements.}
\centering
\footnotesize
\begin{tabular}{ll|l|ll|ll}
& & Baseline & \multicolumn{2}{c}{MPI} & \multicolumn{2}{c}{Erlang} \\
Problem size & & $24$ cores & $192$ cores & $768$ cores & $192$ cores & $768$ cores \\
\hline
$12,288 \times 59 =$ & & & & & &\\
$724,992$ tetrahedra & Time (sec) & $\mathbf{129.81^{*}}$ & $57.74$ & $72.88$ & $97.00$ & $97.00$  \\
& Efficiency & $100.00\%$ & $28.10\%$ & $5.57\%$ & $16.62\%$ & $4.16\%$ \\
$1,536 \times 6,046 =$ & & & & & &\\
$9,286,656$ tetrahedra & Time (sec) & $489.00$ & $\mathbf{82.57^{*}}$ & $51.72$ & $\mathbf{108.00^{*}}$ & $65.00$ \\
& Efficiency & $100.00\%$ & $74.03\%$ & $29.54\%$ & $56.60\%$ & $23.51\%$  \\
$12,288 \times 6,046 =$ & & & & & &\\
$74,293,248$ tetrahedra & Time (sec) & $4,753.00$ & $641.61$ & $\mathbf{220.20^{*}}$ & $667.00$ & $\mathbf{240.00^{*}}$ \\
& Efficiency & $100.00\%$ & $92.60\%$ & $67.45\%$ & $89.70\%$ & $61.89\%$ 
\end{tabular}
\label{tab:measurements}
\end{table*}

The entries in  Table~\ref{tab:measurements} that are in bold and marked with a  `$*$' can be also used for a weak scalability analysis.
We first point out that from the $1^{st}$ to the $3^{rd}$ configuration, using the Erlang version, there was an increase of the time of execution by a factor of $\approx 1.84$, so the weak scalability efficiency was of $54,34\%$.
As for the MPI version, there was an increase of this time by a factor of $\approx 1.69$, with an efficiency of  $58.95\%$. 
We must nevertheless compare these efficiency assessments with 3 different ways to observe the increase in the size of the corresponding problems:
\begin{enumerate} 
\item Taking the total number of tetrahedra in each configuration: from the $1^{st}$ to the $3^{rd}$ configuration there is an increase of $\approx 102.47$ times the number of tetrahedra; 
\item Taking the correspondence with  Galerkin DoFs for $\mathbb{P}_h^{1}$ as presented in Table~\ref{tab:galerkin_corr}: from the $1^{st}$ to the $3^{rd}$ configuration there is an increase of $\approx 61$ times the number of DoFs;
\item Taking the approximation error in the $L^2$-norm of the computed solution: from the $1^{st}$ to the $3^{rd}$ configuration there is a reduction of $\approx 7$ times the error to the exact solution (from $0.151 \times 10^{-2}$ to $0.214 \times 10^{-3}$).
\end{enumerate}

From the numbers above we can verify that the two  simulator prototypes scaled well, with a small increase in the time of execution when compared with the large increase in the size of the problems.

\section{Conclusions}
\label{sec:conc}

In this paper we presented our ongoing work on the implementation of scalable simulators specifically crafted for the MHM methods.
We presented a preliminary evaluation of our two simulator prototypes, whose results are promising with regard to its scalability.
These results also show that Erlang's overhead was small in comparison with its benefits in terms of productivity and extensive support for fault tolerance.

For future work, we will be investigating the use of coprocessors (GPGPUs and MICs) to reduce the time taken by the execution of the large amount of \textsc{solveLocalProblem} primitive instances in the simulators.
We also intend to experiment with other parallel sparse linear solvers than PaStiX to solve Equation~\ref{eq:mat_global} in the \textsc{reduceLocalProblems} primitive for the case of the MPI version. 
In particular, the hybrid approaches presented in~\cite{Agullo201323,yamazaki2013} combine direct and iterative methods and seem fit to solve linear systems similar to the one illustrated in Equation~(\ref{eq:mat_global}).

\section{Acknowledgement}
These research results have received funding from the Bull/France, EU H2020 Programme and from MCTI/RNP-Brazil under the HPC4E Project, grant agreement n° 689772.
The simulations presented herein used the HPC resources provided by LNCC's SDumont supercomputer (\url{http://sdumont.lncc.br}).

%
\label{sect:bib}
\bibliographystyle{plain}


\end{document}